\documentclass[10pt,pra,twoside]{revtex4}
\usepackage[pdftex]{graphicx}
\usepackage{amsmath, amsthm, amssymb}
\usepackage{subfigure}
\usepackage{comment}
\usepackage{url}

\newcommand{\E}{\mathbb{E}}

\newcommand{\ket}[1]{| #1 \rangle}
\newcommand{\bra}[1]{\langle #1|}
\newcommand{\ip}[2]{\langle #1|#2 \rangle}

\newcommand{\be}{\begin{equation}}
\newcommand{\ee}{\end{equation}}
\newcommand{\bea}{\begin{eqnarray}}
\newcommand{\eea}{\end{eqnarray}}
\newcommand{\bes}{\begin{equation*}}
\newcommand{\ees}{\end{equation*}}
\newcommand{\beas}{\begin{eqnarray*}}
\newcommand{\eeas}{\end{eqnarray*}}

\newtheorem{thm}{Theorem}[section]

\begin{document}

\title{A lower bound on entanglement-assisted quantum communication complexity}

\author{Ashley Montanaro}
\affiliation{Department of Computer Science, University of Bristol, Bristol, BS8 1UB, U.K.}
\email{montanar@cs.bris.ac.uk}
\author{Andreas Winter}
\affiliation{Department of Mathematics, University of Bristol, Bristol BS8 1TW, U.K.}
\email{a.j.winter@bristol.ac.uk}

\date{\today}

\begin{abstract}
We prove a general lower bound on the bounded-error entanglement-assisted quantum communication complexity of Boolean functions. The bound is based on the concept that any classical or quantum protocol to evaluate a function on distributed inputs can be turned into a quantum communication protocol. As an application of this bound, we give a very simple proof of the statement that almost all Boolean functions on $n \,+\, n$ bits have linear communication complexity, even in the presence of unlimited entanglement.
\end{abstract}

\maketitle


\section{Introduction}

Consider a total Boolean function $f: \{0,1\}^n \times \{0,1\}^n \mapsto \{0,1\}$. The {\em quantum communication complexity} of $f$ is defined to be the minimum number of qubits required to be transmitted between two parties (Alice and Bob) for them to compute $f(x,y)$ for any two $n$-bit inputs $x$, $y$, given that Alice starts out with $x$ and Bob with $y$. This number is clearly upper bounded by $n$, but for some functions may be considerably lower. Alice and Bob may be allowed some probability of error $\epsilon$, and may be allowed to share an entangled state before they start their protocol. We will assume that Bob has to output the result. (See \cite{dewolf} and \cite{kushilevitz} for excellent introductions to quantum and classical communication complexity, respectively.)

Some functions are known to have a quantum communication complexity lower than their classical communication complexity (for example, a bounded-error protocol for the disjointness function $f(x,y)=1 \Leftrightarrow |x\wedge y|=0$ requires $\Omega(n)$ bits of classical communication, but only $\Theta(\sqrt{n})$ qubits of quantum communication \cite{aaronson,razborov}), but it is still open whether the quantum communication complexity of total functions can ever be exponentially smaller than the classical communication complexity. It is therefore of interest to produce lower bounds on quantum communication complexity. In this context, the model with prior entanglement is less well understood; although there are strong bounds known for some classes of functions \cite{cleve, razborov}, there are few general lower bounds \cite{buhrman}. It has been shown \cite{gavinsky3,gavinsky2} that sharing entanglement may significantly reduce the communication cost of computing a partial function (where there is a promise on the input), but it is unknown whether a similar result may hold for total functions.

In this paper, we develop an elegant result of Cleve et al.\ that relates computation to communication. Cleve et al.\ showed \cite{cleve} that, if Alice and Bob have access to a protocol to exactly compute the inner product function $IP(x,y) = \sum_i x_i y_i$ (mod 2), then this can be used to produce a quantum protocol that communicates Alice's input $x$ to Bob. They used this to show that $IP$ cannot be computed (exactly and without prior entanglement) by sending fewer than $n$ bits from Alice to Bob. Similar results hold for the bounded-error case and with prior entanglement.

We show that a weaker form of this result can be extended to {\em all} Boolean functions. The extension leads to the development of a new complexity measure for Boolean functions: {\em communication capacity}. Given a Boolean function $f(x,y)$, we define the communication capacity of $f$ as the maximum number of bits which the execution of a protocol to compute $f$ allows Alice to communicate to Bob (in an asymptotic sense). This is a concept which has no classical analogue, and which can be shown to give a lower bound on the quantum communication complexity of $f$ (with or without entanglement).

The lower bound we obtain turns out to be a generalisation of a bound obtained by Klauck \cite{klauck} on quantum communication complexity in the model without entanglement. The result here can thus be seen as extending Klauck's bound to the model of entanglement-assisted quantum communication, and giving it a satisfying operational interpretation. As our bound also holds for classical communication complexity, it fits into the framework of results using ideas from quantum information to say something about classical computation.

We will use the standard notation $Q_E(f)$ to denote the quantum communication complexity of $f$ in the case where the protocol must be exact, $Q_\epsilon(f)$ the complexity where Alice and Bob are allowed to err with probability $\epsilon<1/2$, and $Q_2(f)$ the complexity in the case where $\epsilon=1/3$. In all three cases, Alice and Bob's initial state is separable; $Q_E^*(f)$, $Q_\epsilon^*(f)$ and $Q_2^*(f)$ will represent the equivalent quantities in the case where they are allowed to share an arbitrary initial entangled state.

As is usual in computational complexity, we would expect most functions to have ``high'' quantum communication complexity. Kremer showed \cite{kremer} by a counting argument that a random function $f$ has $Q_2(f)\ge n/2$ (and thus $Q_E(f)\ge n/2$). Buhrman and de Wolf extended Kremer's methods to show that, for all $f$, $Q_E^*(f) \ge (\log \mathrm{rank}(f))/2$ \cite{buhrman} (an equivalent result is shown in section 6.4.2 of \cite{nielsen2}). As almost all Boolean matrices have full rank, this shows that for almost all $f$, $Q_E^*(f) \ge n/2$. Very recently, Gavinsky, Kempe and de Wolf \cite{gavinsky} have shown the final remaining case: for almost all $f$, $Q_2^*(f) = \Omega(n)$. Their technique was to relate quantum communication protocols to quantum fingerprinting protocols, and then to show a relationship between quantum fingerprinting and some well-studied concepts from classical computational learning theory. This result was shown independently by Linial and Shraibman \cite{linial}; their paper also extends the well-known discrepancy lower bound to the model of quantum communication with entanglement.

As an application of our communication capacity technique, we reprove the result that for almost all $f$, $Q_2^*(f) = \Omega(n)$. The proof is of a quite different character and of (arguably) a more ``quantum'' nature, as it is based on showing that the entropy of almost all density matrices produced in a certain random way is high.

\subsection{Notation}

We will use $M$ to denote the square communication matrix of $f$ (where $M_{xy}$ is equal to $(-1)^{f(x,y)}$). $H(v)$ will denote the Shannon entropy of a vector $v$, and $S(\rho)$ the von Neumann entropy of a density matrix $\rho$ ($S(\rho)=-\mathrm{tr}(\rho \log \rho)$). All logarithms will be taken to base 2.


\section{Turning any distributed function into a communication protocol}

In this section, we will describe a protocol (which is a simple extension of the protocol in \cite{cleve} for IP) that allows any protocol for evaluating a distributed function to be turned into a communication protocol. However, for some functions, the communication will be considerably more inefficient than IP allows (Alice may only be able to send $\ll n$ bits to Bob).

\subsection{Exact protocols}

Say Alice and Bob have access to a classical or quantum protocol that computes $f(x,y)$ exactly. We express this as a unitary $P$ that performs the following action.
\be P\ket{x}_A\ket{y}_B\ket{0}_B\ket{a}_{AB} = \ket{x}_A\ket{y}_B\ket{f(x,y)}_B\ket{a'}_{AB} \ee
where $\ket{a}$, $\ket{a'}$ are arbitrary (and possibly entangled) ancilla states shared by Alice and Bob. Note that, as $P$ does not modify the first two registers, we may decompose it as follows:
\be P = \sum_{x,y} \ket{x}\bra{x}_A \otimes \ket{y}\bra{y}_B \otimes U_{xy} \ee
for some unitary $U_{xy}$ acting only on the last two registers. Following \cite{cleve}, we will turn this into a ``clean'' protocol $P'$ by giving Bob an additional qubit to copy the answer into, then running the protocol backwards to uncompute the ``junk'' $\ket{a'}$. The steps of the clean protocol are thus
\beas
\mathrm{(i)} && \ket{x}_A\ket{y}_B\ket{0}_B\ket{0}_B\ket{a}_{AB}\\
\mathrm{(ii)} &\rightarrow& \ket{x}_A\ket{y}_B\ket{f(x,y)}_B\ket{0}_B\ket{a'}_{AB}\\
\mathrm{(iii)} &\rightarrow& \ket{x}_A\ket{y}_B\ket{f(x,y)}_B\ket{f(x,y)}_B\ket{a'}_{AB}\\
\mathrm{(iv)} &\rightarrow& \ket{x}_A\ket{y}_B\ket{0}_B\ket{f(x,y)}_B\ket{a}_{AB}
\eeas
where now the fourth register contains the answer. Ignoring the third and fifth registers, which are the same at the beginning and the end of the protocol, we are left with the map
\be P'\ket{x}_A\ket{y}_B\ket{0}_B = \ket{x}_A\ket{y}_B\ket{f(x,y)}_B \ee
Note that, if the original protocol $P$ communicated $a$ qubits from Alice to Bob and $b$ qubits from Bob to Alice, the protocol $P'$ requires $a+b$ qubits to be communicated in each direction. That is, $P'$ sends as many qubits in the ``forward'' direction as the original protocol $P$ sends in total. Now say Alice wants to communicate her input $x$ to Bob using this protocol. They start with the following state, where $(b_y)$ is an arbitrary probability distribution on Bob's inputs:
\be \ket{\psi} = \ket{x}_A \left( \sum_{y\in\{0,1\}^n} \sqrt{b_y} \ket{y}_B \right) \frac{1}{\sqrt{2}}(\ket{0}-\ket{1})_B \ee
Note that this state is separable (so we do not {\em require} entanglement to execute the communication protocol). After executing the clean protocol for $f$, they are left with
\bea
P'\ket{\psi} &=& \ket{x}_A \left( \frac{1}{\sqrt{2}} \sum_{y\in\{0,1\}^n} \sqrt{b_y} \ket{y}_B \left(\ket{f(x,y)}-\ket{1-f(x,y)}\right)_B\right)\\
&=& \ket{x}_A \left( \sum_{y\in\{0,1\}^n} (-1)^{f(x,y)} \sqrt{b_y} \ket{y}_B \right) \frac{1}{\sqrt{2}}\left(\ket{0}-\ket{1}\right)_B
\eea
Ignoring the registers that remain the same throughout, Bob has the following state at the end of the protocol.
\be \ket{\psi_x} = \sum_{y\in\{0,1\}^n} (-1)^{f(x,y)} \sqrt{b_y} \ket{y} \ee
This state provides some information about Alice's bit string $x$. If $\ip{\psi_x}{\psi_{x'}}=0$ for all $x' \neq x$ (as is the case with the protocol of \cite{cleve} for IP, where Bob uses the uniform distribution on his inputs) then Bob can determine $x$ with certainty and hence has received $n$ bits from Alice. If this is not the case, then we can still quantify precisely how much information can be transmitted. The protocol is equivalent to Alice encoding the classical bit-string $x$ as a state $\ket{\psi_x}$, and co-operating with Bob to send it to him. Say Alice uses a distribution $(a_x)$ on her inputs. Then the ensemble representing what Bob eventually receives is
\be \rho=\sum_{x\in\{0,1\}^n} a_x \ket{\psi_x}\bra{\psi_x} \ee
By Holevo's theorem \cite{holevo}, the entropy $S(\rho)$ describes the maximum number of bits of classical information about $x$ available to Bob by measuring $\rho$. And, by the Holevo-Schumacher-Westmoreland channel coding theorem for a channel with pure signal states \cite{hausladen}, Alice and Bob can achieve this bound (in an asymptotic sense) using block coding!

Therefore, the ability to compute $f$ exactly can be used to transmit $S(\rho)$ bits of information through a quantum channel, even though this does not hold if Alice and Bob are restricted to a classical channel. We thus define the {\em communication capacity} of a Boolean function $f$ as the maximum over all probability distributions $(a_x)$ (on Alice's inputs) and $(b_y)$ (on Bob's inputs) of
\be S\left(\sum_{x\in\{0,1\}^n} a_x \ket{\psi_x}\bra{\psi_x}\right) \mbox{~, where~} \ket{\psi_x} = \sum_{y\in\{0,1\}^n} (-1)^{f(x,y)} \sqrt{b_y} \ket{y} \ee


\subsection{Bounded error protocols}

In the case where Alice and Bob have access to a protocol computing $f$ with some probability of error, Bob will not have the state $\ket{\psi_x}$ at the end of the protocol, but rather some approximation $\ket{\psi^\epsilon_x}$. We will now show that, if the error probability is small, this is in fact still sufficient to communicate a significant amount of information from Alice to Bob. As before, Alice will use a distribution $(a_x)$ on her inputs, and Bob a distribution $(b_y)$.

Say Alice and Bob are using a protocol $P^\epsilon$ that computes $f$ with probability of error $\epsilon$, where $\epsilon<1/2$. As before, the $\ket{x}$ and $\ket{y}$ registers will be unchanged by this protocol, so we can write
\be P^\epsilon = \sum_{x,y} \ket{x}\bra{x}_A \otimes \ket{y}\bra{y}_B \otimes U^\epsilon_{xy} \ee
Now let us run the protocol on the same starting state $\ket{\psi}$ as in the previous section.
\beas
\mathrm{(i)} && \ket{x}_A\left(\frac{1}{\sqrt{2}} \sum_{y\in\{0,1\}^n} \sqrt{b_y} \ket{y}_B\ket{0}_B(\ket{0}-\ket{1})_B\right)\ket{a}_{AB} \\
\mathrm{(ii)} &\rightarrow& \ket{x}_A\left(\frac{1}{\sqrt{2}} \sum_{y\in\{0,1\}^n} \sqrt{b_y} \ket{y}_B(\alpha_{xy}\ket{0}+\beta_{xy}\ket{1})_B(\ket{0}-\ket{1})_B\right)\ket{a'}_{AB}
\eeas
where the effect of $U^\epsilon_{xy}$ on the ``answer'' qubit has been decomposed into $\alpha_{xy}$ and $\beta_{xy}$ components. If $f(x,y)=0$, then $|\alpha_{xy}|^2 \ge 1-\epsilon$, and thus (by unitarity) $|\beta_{xy}|^2 \le \epsilon$; if $f(x,y)=1$, $|\beta_{xy}|^2 \ge 1-\epsilon$ and $|\alpha_{xy}|^2 \le \epsilon$. The ancilla register is still completely arbitrary, and in particular may be entangled with any of the other registers. Continuing the protocol, we have

\begin{align}
\mathrm{(iii)} &\rightarrow \ket{x}_A\left(\frac{1}{\sqrt{2}} \sum_{y\in\{0,1\}^n} \sqrt{b_y} \ket{y}_B(\alpha_{xy}\ket{0}\ket{0}-\alpha_{xy}\ket{0}\ket{1} - \beta_{xy}\ket{1}\ket{0} + \beta_{xy}\ket{1}\ket{1})_B\right)\ket{a'}_{AB}\\
\begin{split}
\mathrm{(iv)} &\rightarrow \ket{x}_A\left(\frac{1}{\sqrt{2}} \sum_{y\in\{0,1\}^n} \sqrt{b_y} \ket{y}_B(\alpha_{xy}(\alpha_{xy}^*\ket{0}+\gamma_{xy}^*\ket{1})\ket{0} - \alpha_{xy}(\alpha_{xy}^*\ket{0}+\gamma_{xy}^*\ket{1})\ket{1} \right.\\
& \left.\phantom{\sum_{i=1}^d===========}-\beta_{xy}(\beta_{xy}^*\ket{0}+\delta_{xy}^*\ket{1})\ket{0} + \beta_{xy}(\beta_{xy}^*\ket{0}+\delta_{xy}^*\ket{1})\ket{1})_B\right)\ket{a}_{AB}
\end{split}\\
&= \ket{x}_A\left(\frac{1}{\sqrt{2}} \sum_{y\in\{0,1\}^n} \!\!\!\!\sqrt{b_y} \ket{y}_B((\alpha_{xy} \alpha_{xy}^* - \beta_{xy} \beta_{xy}^*)\ket{0} + (\alpha_{xy} \gamma_{xy}^* - \beta_{xy} \delta_{xy}^*)\ket{1})_B(\ket{0}-\ket{1})_B \right)\!\!\ket{a}_{AB} 
\end{align}
where we introduce $\gamma_{xy}^*$ and $\delta_{xy}^*$ as arbitrary elements of $(U^\epsilon_{xy})^\dag$, subject only to the constraint that $U^\epsilon_{xy}$ be unitary. We may now remove registers that end the protocol unchanged and rewrite Bob's final state as
\be \ket{\psi^\epsilon_x} =  \sum_{y\in\{0,1\}^n} \sqrt{b_y} \ket{y}\left((|\alpha_{xy}|^2 - |\beta_{xy}|^2)\ket{0} + (\alpha_{xy} \gamma_{xy}^* - \beta_{xy} \delta_{xy}^*)\ket{1}\right) \ee
Now, if $f(x,y)=0$, then $|\alpha_{xy}|^2 - |\beta_{xy}|^2\ge 1-2\epsilon>0$, whereas if $f(x,y)=1$, $|\alpha_{xy}|^2 - |\beta_{xy}|^2\le 2\epsilon-1<0$. We may therefore write
\be \ket{\psi^\epsilon_x} = \sum_{y\in\{0,1\}^n} \sqrt{b_y} \ket{y}\left((-1)^{f(x,y)} \cos \theta_{xy} \ket{0} + e^{i\phi_{xy}} \sin \theta_{xy} \ket{1}\right) \ee
where $\theta_{xy}$ is real with $\cos \theta_{xy} \ge 1-2\epsilon$, and $\phi_{xy}$ is an arbitrary phase. Crucially, the form of these states is quite restricted and close to the original $\ket{\psi_x}$. In fact, it is clear that
\be |(\bra{\psi_x}\bra{0})\ket{\psi^\epsilon_x}|^2 \geq (1-2\epsilon)^2 \ee
Set $\rho^\epsilon = \sum_{x\in\{0,1\}^n} a_x \ket{\psi^\epsilon_x}\bra{\psi^\epsilon_x}$. We will compare this to the state $\rho' = \sum_{x\in\{0,1\}^n} a_x \ket{\psi_x}\ket{0}\bra{\psi_x}\bra{0}$ (where of course $S(\rho')=S(\rho)$). We have
\be \|\rho' - \rho^\epsilon\|_1 \leq 2\sqrt{ 1-(1-2\epsilon)^2 } \leq 4\sqrt{\epsilon} \ee
We will use Fannes' inequality \cite{fannes} to show that $S(\rho^\epsilon) \approx S(\rho)$. Define the function
\be \label{eta0} \eta_0(x) = \left\{ \begin{array}{ll} -x \log x & \mbox{for $x \leq 1/e$}\\
1/e \log e & \mbox{for $x > 1/e$} \end{array}\right. \ee
Then Fannes' inequality gives that
\be \label{fannesBound} S(\rho^\epsilon) \ge S(\rho) - 4 \sqrt{\epsilon}n - \log\eta_0(4 \sqrt{\epsilon}) \ee


\subsection{Communication complexity lower bounds from communication capacity}

A lower bound for the communication capacity of a function $f$ can be written down in terms of its communication matrix $M$ as follows. As before, set
\be \rho = \sum_{x\in\{0,1\}^n} a_x \ket{\psi_x}\bra{\psi_x} \mbox{~for~} \ket{\psi_x} = \sum_{y\in\{0,1\}^n} (-1)^{f(x,y)} \sqrt{b_y} \ket{y} \ee
for arbitrary probability distributions $(a_x)$, $(b_y)$ on Alice and Bob's inputs. Define the rescaled Gram matrix $G$ as $G_{ij} = \sqrt{a_i} \sqrt{a_j} \ip{\psi_i}{\psi_j}$. Now it is known \cite{jozsa} that $G$ will have the same eigenvalues as $\rho$, and thus the same entropy. But it can easily be verified that
\be G=(AMB)(AMB)^{\dag} \ee
where $A$ and $B$ are diagonal matrices with $A_{ii}=\sqrt{a_i}$, $B_{ii}=\sqrt{b_i}$. So the eigenvalues of $G$ are simply the singular values squared of $AMB$. We may thus write
\be S(\rho) = H(\sigma^2(AMB)) \ee
where $\sigma^2(M)$ denotes the vector containing the squared singular values of a matrix $M$. We can now produce lower bounds on the quantum communication complexity of $f$ by appealing to the result of Nayak and Salzman \cite{nayak} which states that, if Alice wishes to transmit $n$ bits to Bob over a quantum channel with probability of success $p$, Alice must send $m \ge \frac{1}{2}\left(n - \log \frac{1}{p} \right)$ bits to Bob. If they are not allowed to share prior entanglement, the factor of $1/2$ vanishes. This immediately gives a lower bound on the exact quantum communication complexity of $f$, as lower bounds on the forward communication required for the ``clean'' protocols that we use translate into lower bounds on the total amount of communication needed for any communication protocol.

In the bounded-error case, we can still use the Nayak-Salzman result. Consider a block coding scheme with blocks of length $k$ where each letter $\ket{\psi_x^\epsilon}$ is produced by one use of $f$, as in the previous section. By \cite{hausladen} there exists such a scheme that transmits $k S(\rho^\epsilon)-o(k)$ bits of information with $k$ uses of $f$, as $k\rightarrow \infty$, and probability of success $p \rightarrow 1$. A lower bound on the bounded-error quantum communication complexity of $f$ follows immediately:
\be mk \geq \frac{1}{2} (k S(\rho^\epsilon) - o(k) - o(1) ),\ee
hence, after taking the limit $k \rightarrow \infty$, $p \rightarrow 1$, we find $m \geq \frac{1}{2} S(\rho^\epsilon)$.

In order to reduce the error probability $\epsilon$ to $O(1/n^2)$ (to remove the additive term linear in $n$ in inequality (\ref{fannesBound})), it is sufficient to repeat the original protocol $O(\log n)$ times and take a majority vote \cite{kremer}. Alternatively, using (\ref{fannesBound}) directly gives a better bound for functions for which $S(\rho)$ is linear in $n$. We thus have the following theorem.
\begin{thm}
\label{mainThm}
Let $f:\{0,1\}^n\times \{0,1\}^n \mapsto \{0,1\}$ be a total Boolean function with communication matrix $M$. Then, for any non-negative diagonal matrices $A$ and $B$ with $\|A\|_2=\|B\|_2=1$,
\bea
Q_E(f) &\ge& H(\sigma^2(AMB))\\
Q_E^*(f) &\ge& \frac{1}{2} H(\sigma^2(AMB))\\
Q_\epsilon(f) &\ge& \left\{
\begin{array}{l}
\Omega(H(\sigma^2(AMB))/\log n)\\
H(\sigma^2(AMB)) - 4 \sqrt{\epsilon}n - \log\eta_0(4 \sqrt{\epsilon})
\end{array}
\right.\\
Q_\epsilon^*(f) &\ge& \left\{
\begin{array}{l}
\Omega(H(\sigma^2(AMB))/\log n)\\
\frac{1}{2}(H(\sigma^2(AMB)) - 4 \sqrt{\epsilon}n - \log\eta_0(4 \sqrt{\epsilon}) )
\end{array}
\right.
\eea
where $\eta_0(x)$ is defined as in equation (\ref{eta0}).
\end{thm}
If we use the uniform distribution on Alice and Bob's inputs, then $AMB=M/2^n$. In the case of the models without entanglement, Klauck obtained this specialised result via a different method \cite{klauck}. This theorem can thus be seen as simultaneously extending Klauck's work to the model with entanglement, generalising it, and giving it an operational interpretation. The special case of the uniform distribution was also used by Cleve et al.\ \cite{cleve} to prove their lower bound on the communication complexity of IP.


\section{R\'enyi entropic bounds on communication capacity}

A disadvantage of the von Neumann entropy $S(\rho)$ is the difficulty involved in its computation. The {\em second R\'enyi entropy} $S_2(\rho)$ \cite{renyi} provides an easily computable lower bound on $S(\rho)$. $S_2(\rho)$ is defined as
\be S_2(\rho) = -\log \mathrm{tr}(\rho^2) = -\log \sum_{i,j} |\rho_{ij}|^2 \ee
and we have the fundamental property that $S_2(\rho) \le S(\rho)$. The R\'enyi entropy also obeys the bounds $0 \le S_2(\rho)\le n$. As with the von Neumann entropy, the R\'enyi entropy is a function only of the eigenvalues of $\rho$, so the R\'enyi entropy of the density matrix corresponding to an ensemble of equiprobable states is the same as that of the rescaled Gram matrix corresponding to these states. We can use this to write down a formula for the R\'enyi entropy of a density matrix $\rho$ corresponding to the communication matrix $M$ of a function (as in the previous section, specialising to the uniform distribution on Alice and Bob's inputs), which gives a lower bound on its communication capacity and thus its entanglement-assisted communication complexity.
\bea
S_2(\rho) &=& -\log \mathrm{tr}\left(\frac{1}{2^{4n}} (M M^\dag)^2\right) \\
&=& 4n - \log\left(\sum_{i,j} \left( \sum_k M_{ik} M_{jk} \right)^2 \right) \\
&=& 4n - \log\left(\sum_{i,j,k,l} M_{ik} M_{jk} M_{il} M_{jl} \right)
\eea
R\'enyi entropic arguments have previously been used in a different way by van Dam and Hayden \cite{hayden} to put lower bounds on quantum communication complexity.


\section{The quantum communication complexity of a random function}
In this section, we will show a lower bound on the communication capacity -- and thus the quantum communication complexity -- of a random function (one which takes the value 0 or 1 on each possible input with equal probability). Define the state $\rho$ as
\be \rho = \frac{1}{2^n} \sum_{k\in\{0,1\}^n} \ket{\psi_k}\bra{\psi_k} \mbox{, where~} \ket{\psi_k} = \frac{1}{\sqrt{2^n}} \sum_{i\in\{0,1\}^n} (-1)^{a^k_{i+1}} \ket{i} \ee
where $a^k$ is a randomly generated $2^n$-bit string, and $a^k_i$ represents the $i$'th bit of $a^k$. We will show that the R\'enyi entropy $S_2(\rho)$ is high for almost all $\rho$.

\begin{thm}
$\Pr\left[ S_2(\rho) < (1-\delta)n\right] \le e^{-(2^{\delta n}-1)^2/2}$.
\end{thm}

\begin{proof}
We have
\bea
S_2(\rho) &=& 4n - \log\left(\sum_{i,j} \left( \sum_k M_{ik} M_{jk} \right)^2 \right)\\
&=& 4n - \log\left(\sum_i \left( \sum_k (M_{ik})^2 \right)^2 + \sum_{i\neq j} \left( \sum_k M_{ik} M_{jk} \right)^2 \right)\\
&=& 4n - \log\left(N^3 + T\right)
\eea
where we define $N=2^n$ and $T = \sum_{i\neq j} \left( \sum_k M_{ik} M_{jk} \right)^2$. It is then clear that
\be \Pr\left[ S_2(\rho) < (1-\delta) n \right] = \Pr\left[ T > N^3(N^\delta-1) \right] \ee
Each term in the inner sum in $T$ (the sum over $k$) is independent and picked uniformly at random from $\{-1,1\}$. We will now produce a tail bound for $T$ using ``Bernstein's trick'' (see Appendix A of \cite{alon}): from Markov's inequality we have
\be \label{trick} \Pr\left[ T > a \right] < \E(e^{\lambda T})/e^{\lambda a} < \E(e^{\lambda X_{11}})^{N^2}/e^{\lambda a} \ee
where we define $X_{ij} = \left( \sum_k M_{ik} M_{jk} \right)^2$: each $X_{ij}$ is independent and identically distributed, so $T$ is the sum of $N(N-1)<N^2$ copies of $X_{11}$. It remains to calculate $\E(e^{\lambda X_{11}})$. This can be written out explicitly as follows.
\be \E(e^{\lambda X_{11}}) = \frac{1}{2^N} \sum_{k=0}^N \binom{N}{k} e^{\lambda(N-2k)^2} \ee
It is then straightforward to see (using an inequality from \cite{alon}) that the following series of inequalities holds.
\bea
\E(e^{\lambda X_{11}}) &\le& \frac{1}{2^N} \sum_{k=0}^N \binom{N}{k} \left( e^{\lambda(N-2k)^2} + e^{-\lambda(N-2k)^2} \right) \le \frac{1}{2^{N-1}} \sum_{k=0}^N \binom{N}{k} e^{\lambda^2(N-2k)^4/2}\\
&\le& \frac{1}{2^{N-1}} \sum_{k=0}^N \binom{N}{k} e^{\lambda^2 N^4/2} = 2e^{\lambda^2 N^4/2}
\eea
Inserting this in eqn (\ref{trick}), and minimising over $\lambda$, gives
\be \Pr\left[ T > a \right] < 2e^{-a^2/2N^6} \ee
and substituting $a=N^3(N^\delta-1)$ gives the required result.
\end{proof}
In particular, putting $\delta=1/2$ gives that $\Pr\left[ S_2(\rho) < n/2 \right] \le 2e^{-(\sqrt{N}-1)^2/2}$, which is doubly exponentially small in $n$. As $\rho$ corresponds to the communication matrix of a random function, Theorem \ref{mainThm} immediately gives the result that the entanglement-assisted quantum communication complexity of almost all functions is $\Omega(n)$.


\section{Discussion and open problems}
We have shown that the implementation of any distributed computation between
Alice and Bob entails the ability to communicate from one user to the other.
This communication capacity of a Boolean function of two arguments is
naturally a lower bound on the communication complexity to compute that
function, and we have proved corresponding lower bounds, even in the presence
of arbitrary entanglement.

These bounds show that random functions of two $n$-bit strings mostly have communication
complexity close to $n$. However, in general it has to be noted that our bounds
are not that good: an example is provided by the set-disjointness problem,
where Alice and Bob want to determine if their strings $x$ and $y$ have
a position where they are both $1$. It is known that the quantum communication
complexity of this function is $\Theta(\sqrt{n})$~\cite{razborov,aaronson}.
On the other hand, the entropy in our main theorem was already computed for this case
in~\cite{ambainis}, and it is only $O(\log n)$. Thus, not quite
surprisingly, the ability of a function to let Alice communicate to Bob
is not the same as the communication cost of implementing this computation.

Looking again at our main theorem, we are left with one interesting question:
is the logarithmic factor that we lose in the bounded error model really
necessary? It appears to be a technicality, since we need to boost the success
probability to apply Fannes' inequality, but we were unable to determine
if it is just that or if there are cases in which the lower bound is tight.


\section*{Acknowledgements}
AM would like to thank Richard Jozsa for careful reading and comments on this manuscript, and Tony Short and Aram Harrow for helpful discussions. We thank Ronald de Wolf for pointing out references \cite{linial} and \cite{nielsen2}. AW acknowledges support via the EC project QAP, as well as from the U.K. EPSRC. He also gratefully notes the hospitality of the Perimeter Institute for Theoretical Physics in Waterloo, Ontario, where part of this work was done.


\end{document}